\documentclass{PoS}
\usepackage{graphicx}
\usepackage{subfigure}
\usepackage{url}
\usepackage{natbib}

\title{Radio Observations of Star Forming Galaxies in the SKA era}

\ShortTitle{Radio counts of star-forming galaxies}

\author{
\speaker{Claudia Mancuso}$^1$,
Andrea Lapi$^{1,2}$,
Zhen-Yi Cai$^{3}$,
Mattia Negrello$^{4}$,
Gianfranco De Zotti$^{1,4}$,
Francesca Perrotta$^{1}$,
Luigi Danese$^{1}$
\\
$^1$Astrophysics Sector, SISSA, Via Bonomea 265, I-34136 Trieste ;
$^2$Dipartimento di Fisica, Universit\`a `Tor Vergata', Via della Ricerca Scientifica 1, I-00133 Roma, Italy ;
$^3$Center for Astrophysics, University of Science and Technology of China, Hefei, 230026, China;
$^4$INAF - Osservatorio Astronomico di Padova, Vicolo dell'Osservatorio 5, I-35122 Padova, Italy
\\
E-mail: \email{cmancuso@sissa.it}}

\abstract{We have combined determinations of the epoch-dependent star formation rate
(SFR) function with relationships between SFR and radio (synchrotron and
free-free) emission to work out detailed predictions for the counts and the
redshift distributions of star-forming galaxies detected by planned Square
Kilometer Array (SKA) surveys. The evolving SFR function comes from recent
models fitting the far-infrared (FIR) to millimeter-wave luminosity functions
and the ultraviolet (UV) luminosity functions up to $z=10$,  extended to take
into account additional UV survey data. We used very deep 1.4\,GHz number
counts from the literature to check the relationship between SFR and
synchrotron emission, and the 95\,GHz South Pole Telescope (SPT) counts of
dusty galaxies to test the relationship between SFR and free-free emission.
We show that the SKA will allow us to investigate the SFRs of galaxies down
to few $M_\odot/\hbox{yr}$ up to z=10, thus extending by more than two orders
of magnitude the high-$z$ SFR functions derived from \textit{Herschel}
surveys. SKA1-MID surveys, down to $\mu$Jy levels, will detect hundreds of
strongly lensed galaxies per square degree; a substantial fraction of them
will show at least two images above the detection limits.}

\FullConference{Advancing Astrophysics with the Square Kilometre Array\\
June 8-13, 2014\\ Giardini Naxos, Italy}

\newcommand{\skipthis}[1]{}

\newcommand\apj{ApJ}

\newcommand\mnras{MNRAS}

\newcommand\aap{A$\&$A}
\newcommand\araa{ARA$\&$A}
\newcommand\aj{AJ}
\newcommand\pasa{PASA}
\newcommand\physrep{PhR}

\begin{document}

\section{Introduction}\label{sect:intro}

The huge amount of infrared (IR) to millimeter-wave data that has been
accumulating in the last several years thanks to \textit{Spitzer},
\textit{Herschel}, SCUBA and South Pole Telescope (SPT) surveys has made
possible spectacular advances in our understanding of galaxy evolution. In
fact the interstellar dust absorbs and re-emits at IR wavelengths about half
of the starlight of the Universe  \citep{Casey2014}. Hence, the evolution of
the IR luminosity function directly maps that of the dust-enshrouded star
formation rate (SFR). \textit{Herschel} data are particularly powerful in
this respect as they probe the dust emission peak, thus providing the best
estimates of the total IR luminosity.

However, the IR emission misses the starlight not absorbed by dust and
therefore underestimates the SFR whenever the absorption optical depth is not
very high. This is the case for the earliest phases of galaxy evolution when
the metal enrichment of the interstellar medium is just beginning, as well as
for dust-poor dwarf galaxies and metal-poor regions of more-massive galaxies
\citep{KennicuttEvans2012}. A complete inventory of the SFR requires a
combination of IR and UV photometry and thus suffers from limitations in both
wavebands.

An important alternative tool for measuring the cosmic star formation history
of the Universe is provided by deep radio continuum surveys. A tight
relationship between low-frequency radio (synchrotron) and IR luminosity
(hence SFR) has long been established
\citep{Helou1985,Condon1992,Yun2001,Ivison2010,Jarvis2010,Bourne2011,Mao2011}.
However the physical basis of this relationship is not yet totally clear. In
fact many physical processes (propagation of relativistic electrons, strength
and structure of the magnetic field, size and composition of dust grains)
must conspire together to produce this relation \citep{Bell2003, Helou1993,
Niklas1997, Murphy2009,Lacki2010,Hippelein2003}. Hence it is not granted that
the relation also applies to redshift/luminosity ranges where the available
data are insufficient to test it accurately. Moreover we cannot be sure that
the observed synchrotron emission is not contaminated by faint nuclear
activity.

There is a second process contributing to the radio emission of star-forming
galaxies: the free-free emission from hot electrons, which is directly
proportional to the production rate of ionising photon  by young, massive
stars. It shows up at rest frame frequencies of tens of GHz, where it is
generally optically thin, and thus offers a clean way to quantify the current
star formation activity in galaxies. This picture could be complicated by the
presence of anomalous dust emission \citep[][and references
therein]{PlanckCollaborationXX2011} which occurs at similar frequencies and
is thought to arise from spinning dust grains
\citep[e.g.,][]{DraineLazarian1998}. However there is currently no evidence
that this component contributes significantly to globally integrated
measurements \citep{Murphy2012}. Thus high frequency radio observations may
be particularly powerful for precisely measuring the star formation history
of the Universe.

Very deep radio surveys have shown that, at GHz frequencies, the counts below
100--$200\,\mu$Jy are dominated by star-forming galaxies \citep[][and
references therein]{Padovani2011}. Current surveys only extend to a few tens
of $\mu$Jy, i.e. cover a flux density range where, at low radio frequencies,
the detected radio emission is of synchrotron origin. Only with the advent of
the Square Kilometer Array (SKA) we expect that the high-$z$ star-forming
galaxies can be seen via their free-free emission (see also Murphy, E. J., et
al. 2015, "The Astrophysics of Star Formation Across Cosmic Time at z $\geq$
10 GHz with the Square Kilometre Array", in proceedings of "Advancing
Astrophysics with the Square Kilometre Array", \pos{PoS (AASKA14)085}).

In this paper we carry out a thorough investigation of the radio counts of
star forming galaxies. In Sect.~\ref{sect:model} we present a short outline
of the model, that builds on the work by \citet{Cai2013} and \citet{Cai2014}.
In Sect.~\ref{sect:calibration} we discuss the calibration of the relation
between radio emissions and SFR. The South Pole Telescope (SPT) surveys at mm
wavelengths are especially useful to test the SFR/free-free relation; this
motivated a re-analysis of the SPT 95 GHz sample of dusty galaxies. These
relations allow us to extend to radio frequencies the \citet{Cai2013} model
for the cosmological evolution of star forming galaxies and exploit it, in
Sect.~\ref{sect:SKAcounts}, to work out predictions for the counts of such
galaxies in the range 1.4 -- 30 GHz. We also compare the coverage of the
SFR--$z$ plane by \textit{Herschel}, UV surveys and SKA. Finally,
Sect.~\ref{sect:conclusions} summarizes our main conclusions.

Throughout this paper we adopt a flat $\Lambda \rm CDM$ cosmology with matter
density $\Omega_{\rm m} = 0.32$, $\Omega_{\rm b} = 0.049$, $\Omega_{\Lambda}
= 0.68$, Hubble constant $h=H_0/100\, \rm km\,s^{-1}\,Mpc^{-1} = 0.67$,
spectrum of primordial perturbations with index $n = 0.96$ and normalization
$\sigma_8 = 0.83$ \citep{PlanckCollaborationXVI2013}.

\section{Outline of the model}\label{sect:model}

A direct tracer of recent star formation is the UV emission of galaxies,
coming from the photospheric emission of massive young stars. In recent years
a great effort has been made  to measure the UV luminosity functions up to
high redshifts
\citep{Bouwens2008,Bouwens2011,Smit2012,Oesch2012,Oesch2013a,Oesch2013b,Schenker2013,McLure2013},
with the aim of reconstructing the history of cosmic re-ionization.

However, as the chemical enrichment of the ISM proceeds and, correspondingly,
the dust abundance increases, a larger and larger fraction of starlight is
absorbed and re-emitted at far-IR wavelengths. The most active star-formation
phases of high-$z$ galaxies indeed suffer by strong dust obscuration and are
most effectively studied in the far-IR/sub-mm region.

A comprehensive investigation of the evolution of the IR luminosity functions
has been recently carried out by \citet{Cai2013} based on a ``hybrid''
approach that reflects the observed dichotomy in the ages of stellar
populations of early-type galaxies on one side and late-type galaxies on the
other \citep[cf.][ their Fig.~10]{Bernardi2010}. Early-type galaxies and
massive bulges of Sa galaxies are composed of relatively old stellar
populations with mass-weighted ages $\gtrsim 8$--9\,Gyr (corresponding to
formation redshifts $z\gtrsim 1$--1.5), while the disk components of spirals
and the irregular galaxies are characterized by significantly younger stellar
populations. Thus the progenitors of early-type galaxies, referred to as
proto-spheroidal galaxies or protospheroids, are the dominant star-forming
population at $z\gtrsim 1.5$, while IR galaxies at $z\lesssim 1.5$ are mostly
late-type ``cold'' (normal) and ``warm'' (starburst) galaxies.

The \citet{Cai2013} model accurately fits a broad variety of
data\footnote{See figures in
\url{http://people.sissa.it/~zcai/galaxy_agn/index.html}.}: multi-frequency
and multi-epoch luminosity functions of galaxies and AGNs, redshift
distributions, number counts (total and per redshift bins). Moreover, it
accurately accounts for the recently determined counts and redshift
distribution of strongly lensed galaxies detected by the South Pole Telescope
\citep[SPT;][]{Mocanu2013, Weiss2013}, published after the paper was
completed \citep{Bonato2014}.

In general the total (8--$1000\,\mu$m) IR luminosity, $L_{\rm IR}$, is a good
proxy of the obscured SFR when the dust heating is dominated by young stars.
In disks of normal galaxies, however, the IR luminosity is the sum of a
``warm'' component heated by young stars and of a ``cold'' (or ``cirrus'')
component, heated by the general radiation field that may be dominated by
older stars. This issue was investigated by \citet{Clemens2013} using a
complete sample of local star-forming galaxies detected by \textit{Planck}.
We adopt the relation between SFR and $L_{\rm IR}$ derived by these authors,
for a Chabrier initial mass function (IMF).

As mentioned in Sect.~\ref{sect:intro}, to get a complete census of the
cosmic SFR we need to complement IR measurements with UV SFR tracers to
measure the unattenuated starlight. For the high-$z$ proto-spheroidal
galaxies we adopted the accurately tested physical model for the evolution of
the UV luminosity function  worked out by \citet{Cai2014} in the framework of
the scenario proposed by \citet{Granato2004} and further elaborated by
\citet{Lapi2006,Lapi2011}, \citet{Mao2007}, and \citet{Cai2013}.

For the low-$z$ galaxies we have supplemented the \citet{Cai2013} model for
evolution of late-type galaxies in the IR with a parametric model for
evolution in the UV. Briefly, the UV ($\lambda = 1500\,$\AA) luminosity
function is described by:
\begin{equation}\label{eq:LF}
	\Phi (\log L_{1500}, z) \hbox{d}\,\log L_{1500} = \Phi^* \Big(\frac{L_{1500}}{L^*}\Big)^{1-\alpha}
	 \times \exp \Big[- \frac{\log ^2 (1+L_{1500}/L^*)}{2\sigma^2} \Big] \hbox{d}\,\log L_{1500}.
\end{equation}
A simple pure luminosity evolution model ($L^*(z) = L^*_0 (1+z)^{\alpha_L}$
up to $z = 1$) turned out to provide a sufficiently good description of the
data (Mancuso et al., in preparation). The best-fit values of the parameters
are $\log(\Phi^*_0/[\rm dex^{-1}\,Mpc^{-3}]) = -2.150 \pm 0.095$,
$\log(L^*_0/L_\odot) = 9.436 \pm 0.119$, $\alpha = 1.477 \pm 0.050$, $\sigma
= 0.326 \pm 0.035$, and $\alpha_L = 2.025 \pm 0.063$.

The \citet{KennicuttEvans2012} calibrations were adopted to convert the UV
luminosity functions into SFR functions. The total redshift-dependent SFR
functions were then computed summing the IR-based and the UV-based ones.

\begin{figure}
\makebox[\textwidth][c]
{
\includegraphics[width=0.5\textwidth, angle=0]{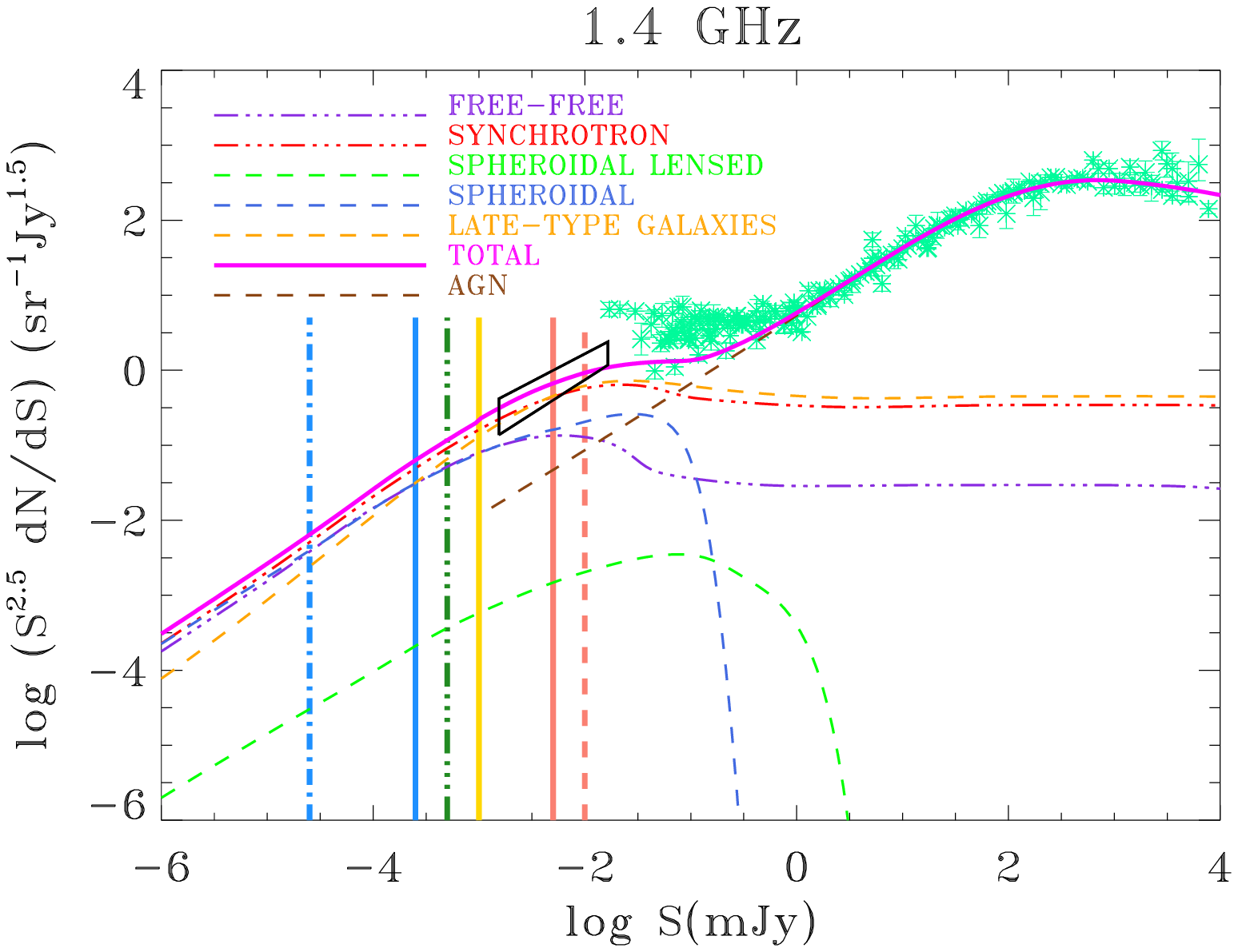}
\includegraphics[width=0.5\textwidth, angle=0]{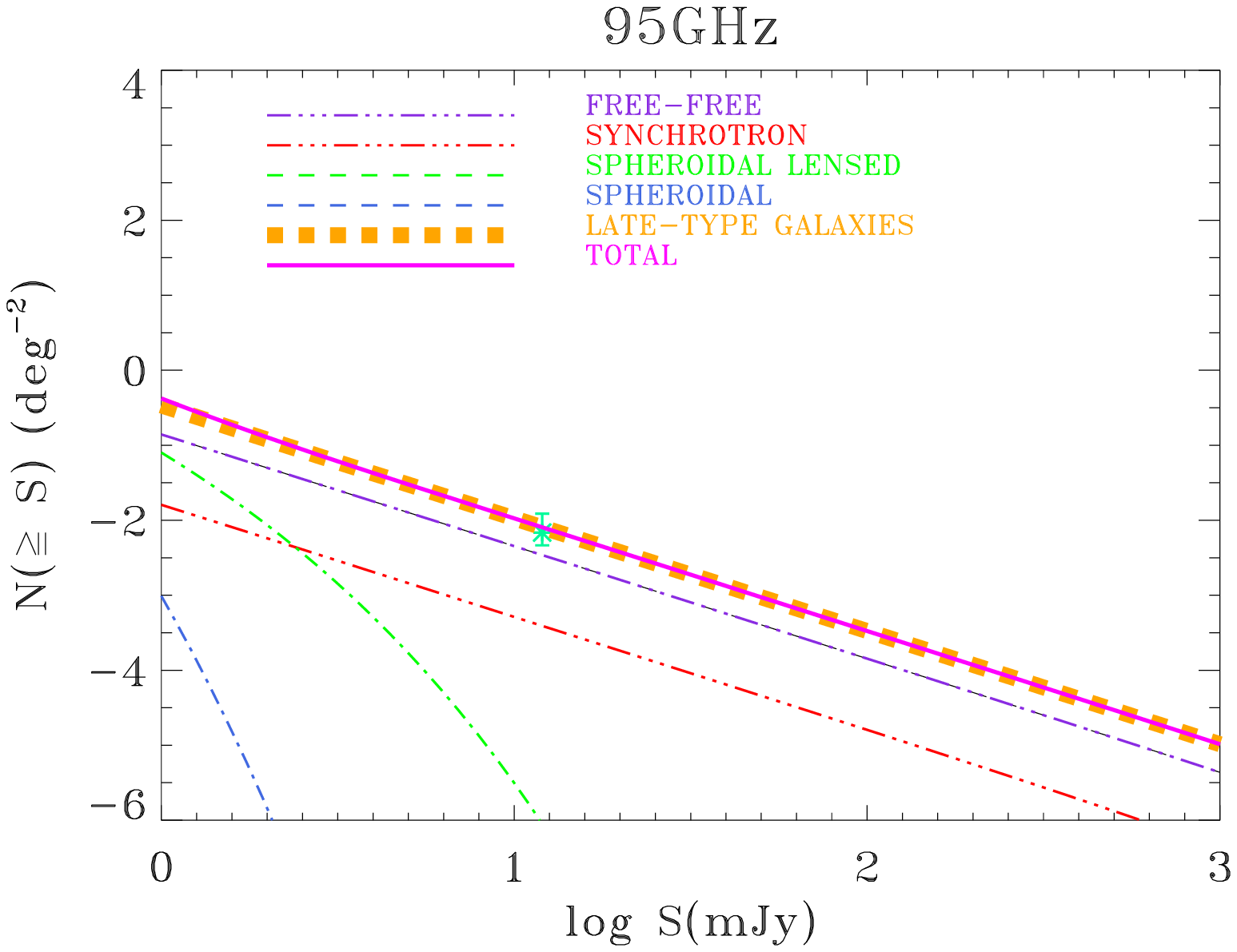}
}
\vspace{-0.8cm}
\caption{\emph{left panel}: Euclidean normalized differential counts at 1.4 GHz compared with model predictions. The dashed brown line shows the best fit model for radio AGNs by \citet{Massardi2010}. The other dashed lines show the contributions of the star-forming populations considered in this paper. The two triple-dot dashed lines show the total synchrotron and free-free emissions from these populations. The solid vertical lines correspond, from left to right, to three $5\,\sigma$ detection limits of preliminarily planned SKA1-MID surveys. The vertical blue dash-dotted line and the pink dashed line are respectively the flux limits for Ultra-Deep SKA2 (MID) and for the 50$\%$ Wide SKA1-MID  surveys, while the green dash-dot-dotted line is the limit for both 50$\%$ Deep SKA1-MID and Wide SKA2 (MID)  surveys.  \emph{Right panel}: our re-assessment of the integral counts of dusty galaxies at $\nu$=95 GHz (asterisks with error bars) compared with the expectations from the \citet{Cai2013} model.
}\label{fig:1.4_95GHz_counts}
\end{figure}

\section{Calibration of the relation between radio emission and SFR}\label{sect:calibration}

A tight linear correlation between the 1.4 GHz luminosity, dominated by
synchrotron, and the IR luminosity has been established since many years
\citep{Condon1992}. A calibration of the relation between the SFR and the
synchrotron emission was calculated by \citet{Murphy2011}. In order to take
into account  electron ageing effects \citep{BandayWolfendale1991} we have
adopted a steepening by $\Delta \alpha=0.5$ above a break frequency of 20
GHz.  The relationship between synchrotron luminosity and SFR then writes:
\begin{equation}\label{eq:Lsync}
L_{\rm sync}\simeq 1.9\times 10^{28} \left(\frac{\hbox{SFR}}{\hbox{M}_{\odot}\hbox{yr}^{-1}}\right) \left(\frac{\nu}{\hbox{GHz}}\right)^{-0.85}\left[1+\left(\frac{\nu}{ 20\rm GHz}\right)^{0.5}\right]^{-1}\, \hbox{erg}\,\hbox{s}^{-1}\,\hbox{Hz}^{-1}.
\end{equation}
Coupling this relation with the redshift dependent SFR functions yielded by
the model outlined in Sect.~\ref{sect:model} we get a good fit to the sub-mJy
1.4 GHz counts (Fig.~\ref{fig:1.4_95GHz_counts}, left panel) without any
adjustment of the parameters. The high-frequency synchrotron emission is
increasingly suppressed with increasing $z$,  as the timescale for energy
losses of relativistic electrons by inverse Compton scattering off the Cosmic
Microwave Background photons decreases as $(1+z)^{-4}$ \citep{Norris2013,
Carilli2008, Murphy2009}. This may lower the counts at the faintest flux
densities, but not by a large factor since, as illustrated by
Fig.~\ref{fig:1.4_95GHz_counts} (left panel), below the limits of current
surveys the free-free contribution is comparable to the synchrotron one.

A relationship between SFR and free-free emission was derived by
\citet{Murphy2012}. We have reformulated it as:
\begin{equation} \label{eq:Lff}
 L_{\rm ff}=3.75\times 10^{26} \left(\frac{\hbox{SFR}}{M_\odot/\hbox{yr}}\right) \, \left(\frac{T}{10^4\,\hbox{K}}\right)^{-0.5}\, \hbox{g}(\nu,\hbox{T})\,\exp{\left(-\frac{h\nu}{ k\hbox{T}}\right)}\, \hspace{3pt} \hbox{erg}\,\hbox{s}^{-1}\,\hbox{Hz}^{-1}
 \end{equation}
where T is the temperature of the emitting plasma and
$\hbox{g}(\nu,\hbox{T})$ is the Gaunt factor for which we adopt the
approximation proposed by \citet{Draine2011} which is more accurate than the
one used by  \citet{Murphy2012}. The coefficient of eq.~(\ref{eq:Lff}) was
computed requiring that this equation equals that  by  \citet{Murphy2012} for
$\nu=33\,$GHz (the frequency at which the relation was calibrated), $T=
10^4\,$K and a pure hydrogen plasma.

We note that the calibrations of the above relationships are based on the
Kroupa IMF while that between the IR emission and the SFR
(Sect.~\ref{sect:model}) relies on the Chabrier IMF. However, as shown by
\citet{ChomiukPovich2011} the two IMFs give almost identical calibrations.

To test the $L_{\rm ff}$-SFR relation we used the South Pole Telescope (SPT)
observations of dusty galaxies at 95\,GHz \citep{Mocanu2013}, since we expect
that, at this frequency, the free-free emission shows up clearly in local
galaxies not hosting a radio loud AGN. To estimate the 95\,GHz counts of
dusty galaxies \citet{Mocanu2013} adopted a statistical approach. They choose
the local minimum in the distribution of the $\alpha^{150}_{220}$ spectral
indices,  $\alpha^{150}_{220}=1.5$, as the threshold for source
classification and computed, for each source, the probability that their
posterior $\alpha^{150}_{220}$ is greater than the threshold value,
$P(\alpha^{150}_{220} > 1.5)$. This quantity was interpreted as the
probability that a source is dust-dominated. The $95$ GHz differential counts
were computed as the sum of probabilities $P(\alpha^{150}_{220} > 1.5)$.
Since the fraction of dusty galaxies is much lower than that of synchrotron
dominated sources, this statistical approach is endowed with large
uncertainties and may strongly overestimate the counts of dusty galaxies.

We have then re-estimated the counts using a lengthier but safer approach,
i.e. we have checked the SED of each  95\,GHz source brighter than the 95\%
completeness limit of 12.6 mJy, collecting all the photometric data available
in the literature.  To model the SEDs we considered both synchrotron and
free-free emission, as well as thermal dust emission. We found that only 4 sources, all with $P(\alpha^{150}_{220} >
1.5)\simeq 1$, are indeed dusty galaxies. This is a factor $\simeq 3$ below
the number found by \citet{Mocanu2013}. The corresponding integral count is
shown in the right panel of Fig.~\ref{fig:1.4_95GHz_counts}.

\begin{figure*}
\hspace{+0.0cm}
\centering1
{
\includegraphics[width=0.764\textwidth, angle=0]{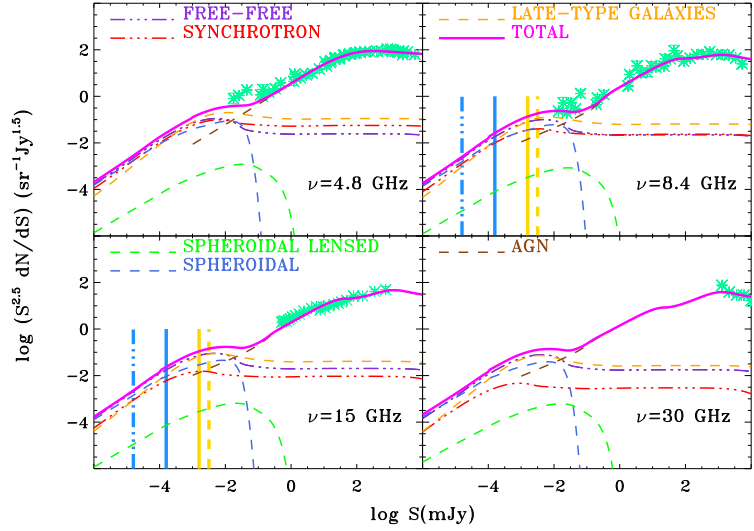}
}
\vspace{-0.8cm}
\caption{Predicted versus observed counts at 4.8, 8.4, 15 and 30 GHz. Dusty galaxies come up at sub-mJy flux density levels and their counts are accounted for by the model. At higher flux densities the counts are dominated by canonical, AGN powered radio sources; the models shown are from \citet{Massardi2010} at 4.8 GHz and from \citet{DeZotti2005} at 8.4, 15 and 30 GHz. The solid vertical lines show the limits for predicted deep (yellow) and ultra deep (blue) band 5 SKA1-MID surveys. The dash-dotted blue line and the dashed yellow line are the limits for an Ultra Deep SKA2 (MID) survey and a 50$\%$ SKA1-MID Deep survey respectively.}
\label{fig:RadioCounts}
\end{figure*}

Further tests and/or predictions of the model are provided by counts at 4.8,
8.4 ,15 and 30 GHz (Fig.~\ref{fig:RadioCounts}). The relative importance of
free-free compared to synchrotron obviously increases with increasing
frequency. While at 1.4 GHz the free-free contribution is always below the
synchrotron one, at 8.4 GHz it takes over at tens of $\mu$Jy levels.

\begin{figure}
\hspace{+0.0cm}
\centering
{
\includegraphics[width=1.0\textwidth, angle=0]{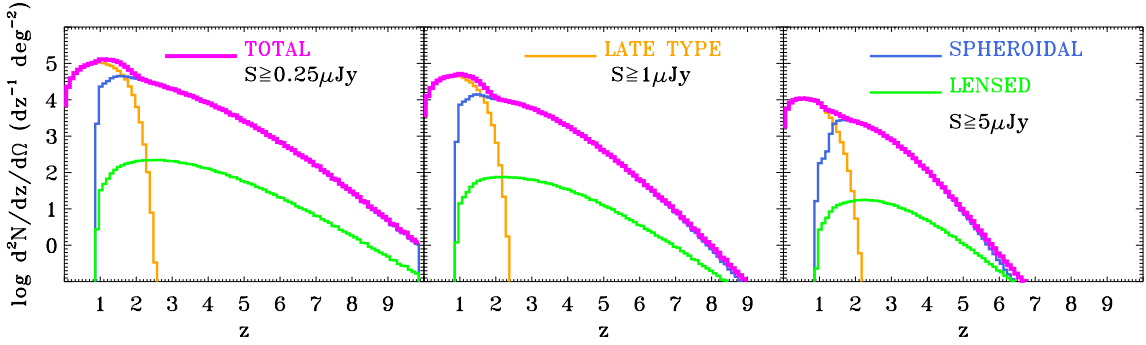}
}
\vspace{-0.5cm}
\caption{Predicted redshift distributions for 1.4 GHz surveys down to 0.25, 1 and $5\,\mu$Jy, which are the $5\,\sigma$ flux density limits of preliminarily planned SKA1-MID surveys.}
 \label{fig:RedshiftDistr}
\end{figure}
%

\begin{figure}
\centering
\includegraphics[width=0.44\textwidth, angle=0]{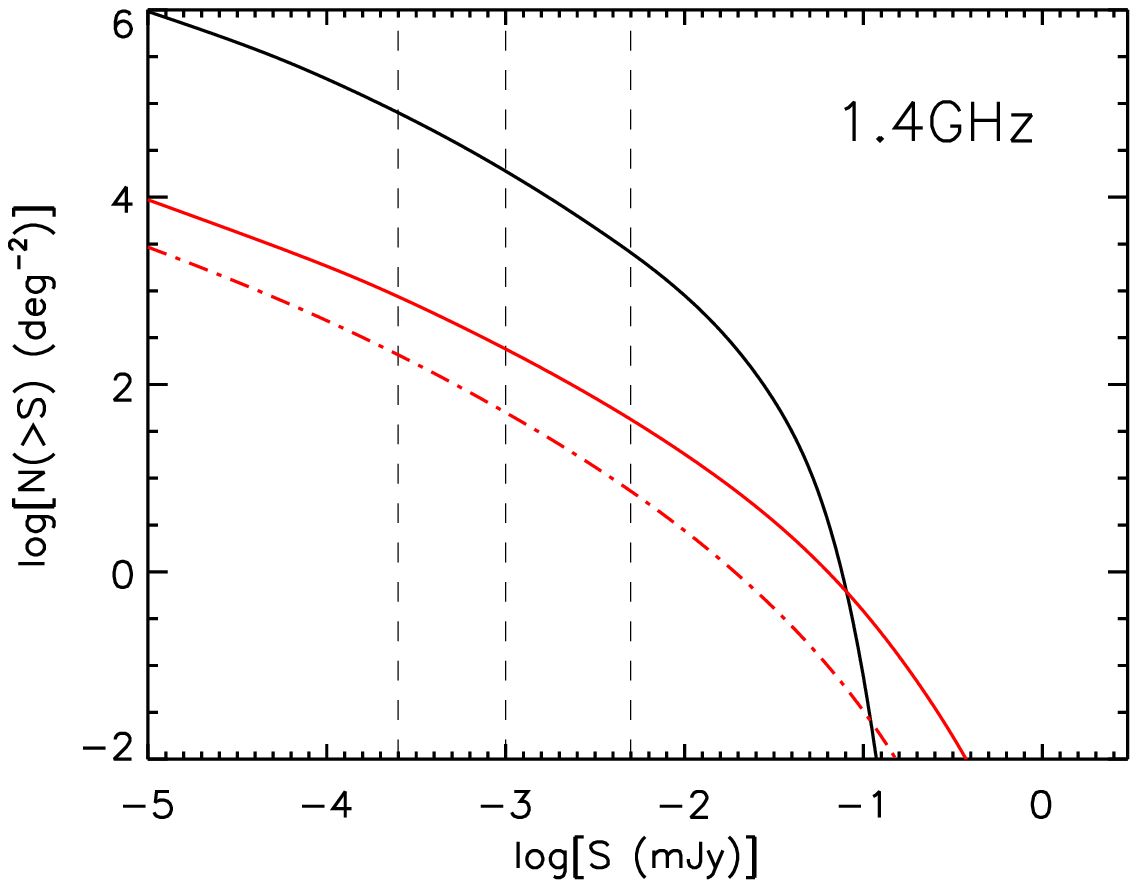}
\includegraphics[width=0.53\textwidth, angle=0]{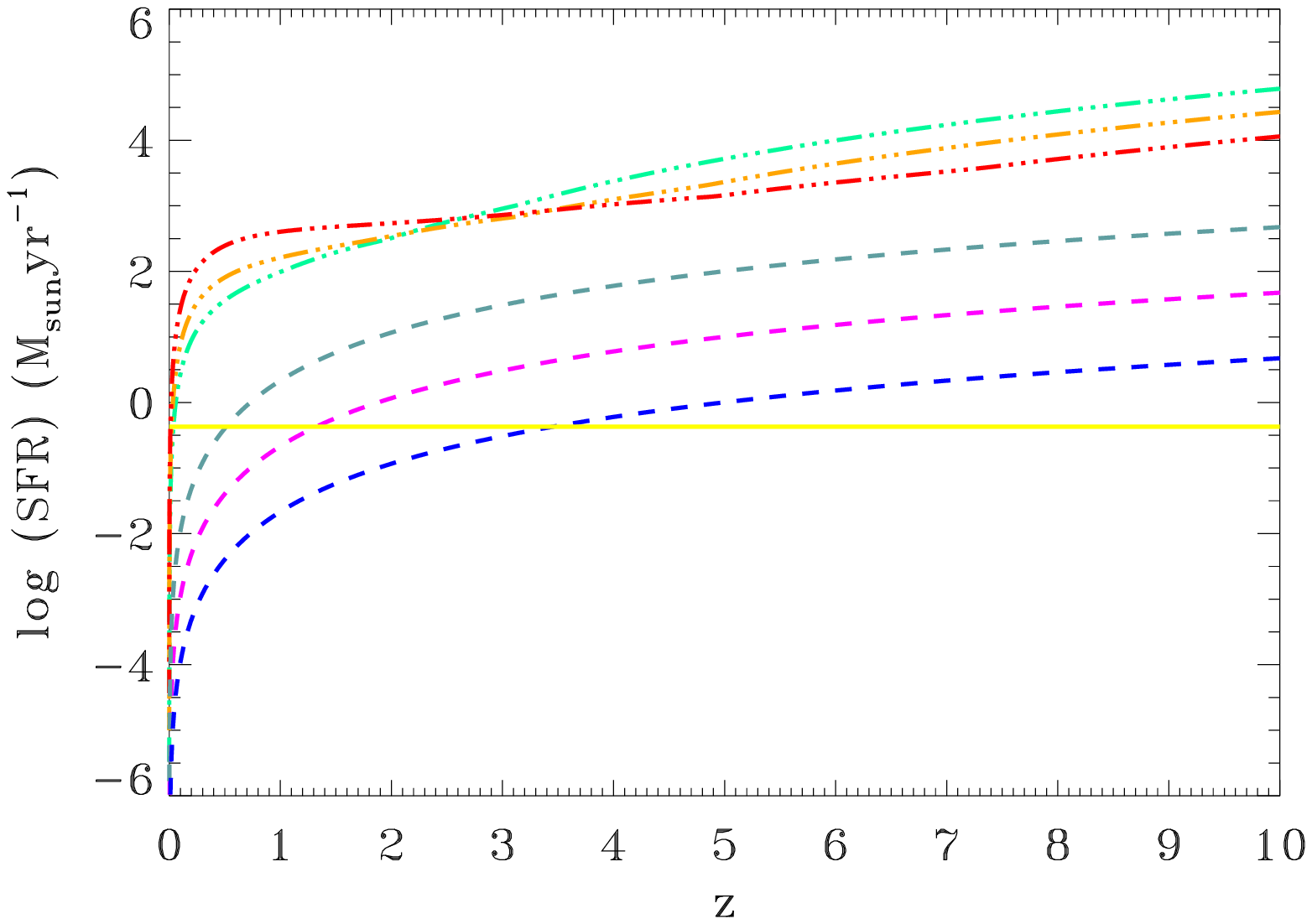}
\vspace{-0.3cm}
\caption{\textit{left panel:} integral number counts of proto-spheroids at 1.4 GHz. Solid black line: unlensed proto-spheroids; solid red line: strongly lensed protospheroids as a function of the total flux density (sum of all images of a single source); dot-dashed red line: counts of strongly lensed proto-spheroids as a function of the flux density of the second brightest image. The vertical dashed lines mark the flux density limits of the possible surveys for SKA1-MID. \textit{Right panel:} minimum SFR detectable by SKA1-MID (dashed magenta line), 50$\%$ SKA1-MID (dashed light blue line) and  SKA2 (MID) (dashed blue line) Ultra-Deep surveys, as a function of $z$, compared with the minimum SFR detected by \textit{Herschel} surveys (dashed-dot-dot-dot lines, green for 250 $\mu$m, orange for 350 $\mu$m and red for 500 $\mu$m) and by UV/H$\alpha$ surveys (horizontal yellow line; see text).}
 \label{fig:lensing_negrello}
\end{figure}

\section{Predictions for surveys with the Square Kilometer Array}\label{sect:SKAcounts}

Preliminary plans for the phase\,1 SKA-MID include a set of surveys at $\sim$
1-1.4 GHz aimed at investigating the galaxy evolution: an Ultra-Deep survey
over 1 deg$^2$ with rms $\sim$ 50 nJy/beam, a Deep survey over 10-30 deg$^2$
with rms $\sim$ 0.2 $\mu$Jy/beam, and a Wide survey over 1000-5000 deg$^2$
with rms $\sim$ 1 $\mu$Jy/beam (see \citet{Dewdney2013}, \citet{Braun2013}).
The corresponding $5\,\sigma$ limits are indicated by vertical solid lines in
the left panel of Fig.~\ref{fig:1.4_95GHz_counts}; the other vertical lines
indicate the detection limits achievable with 50$\%$ of  SKA1-MID Ultra-Deep
and Wide surveys and (full) SKA2 (MID) Ultra-Deep and Wide surveys
sensitivity. We also show the contributions to the 1.4 GHz Euclidean
normalized differential number counts of the three populations of dusty
galaxies considered by \citet{Cai2013}. The main contributors to the ``bump''
at tens of $\mu$Jy levels are late-type galaxies at $z\simeq 1$--1.5. Higher
$z$ proto-spheroidal galaxies become increasingly important at lower flux
densities, down to a few hundred nJy's.

The predicted redshift distributions for surveys at the SKA1-MID flux density
limits are shown in Fig.~\ref{fig:RedshiftDistr}. The fraction of galaxies at
very high redshifts ($z\ge 6$) increases rapidly with decreasing flux
density. At a few hundred nJy levels we expect detections of galaxies at $z$
of up to 10, making possible to investigate the cosmic SFR across the
re-ionization epoch. Note that, although the deepest surveys with the HST are
getting close to that, they inevitably miss the dust-obscured star formation,
while dust obscuration does not affect SKA measurements.

The fraction of strongly lensed galaxies in flux-limited surveys increases
with redshift, to the point that these sources are dominant at the highest
redshifts (see Fig.~\ref{fig:RedshiftDistr}). The predicted counts for
magnifications $\mu \ge 2$ are illustrated in the left panel of
Fig.~\ref{fig:lensing_negrello}, where the solid black line represents the
unlensed proto-spheroidal galaxies, while the red lines represent the lensed
galaxies as a function of the total flux density (solid) or of the flux
density of the second brightest image (dot-dashed). The model predicts 655,
204 and 40 strongly lensed galaxies per deg$^2$ brighter than $0.25$, 1 and
$5\,\mu$Jy, respectively; for 250, 100 and 20 of them, respectively, the
SKA1-MID will directly detect at least two images. A more detailed discussion
on the gravitational lens statistics with SKA is presented in  McKean, J., et
al. 2015, "Strong Gravitational Lensing with the SKA", in proceedings of
"Advancing Astrophysics with the Square Kilometre Array", \pos{PoS
(AASKA14)084}.

The right panel of Fig.~\ref{fig:lensing_negrello} compares the SKA potential
in measuring the evolution of the cosmic SFR to the outcome of
\textit{Herschel} and of the deepest UV and H$\alpha$ surveys. The minimum
luminosities, hence the minimum SFRs, reached by the latter surveys vary
little with redshift. The yellow horizontal line corresponds to their
average. Planned SKA surveys can detect galaxies with SFRs from tens to
hundred $M_\odot/$yr, up to the highest redshifts, extending the SFR
functions by up to 3 orders of magnitude compared with \textit{Herschel},
thus encompassing SFRs typical of $L_\star$ galaxies.

\section{Conclusions}\label{sect:conclusions}

We have worked out detailed predictions of the counts and redshift
distributions for planned SKA surveys, distinguishing the contributions of
the different populations of star-forming galaxies: normal late-type,
starburst and proto-spheroidal galaxies. The predictions are based on models
by \citet{Cai2013} and \citet{Cai2014}, that fit a broad variety of UV and
far-IR/sub-mm data relevant to determine the epoch-dependent SFR function.
These models, however, do not include the contribution to the SFR functions
of moderate to low redshift late-type and starburst galaxies. We have
upgraded them adding these populations. The upgraded models were combined
with the relationships between SFR and radio (synchrotron and free-free)
emission derived by \citet{Murphy2011,Murphy2012}. Such relationships has
been checked exploiting the deepest 1.4\,GHz counts and with data from the
95\,GHz SPT survey, that we have re-analyzed finding that the published SPT
counts of dusty galaxies are overestimated by a factor $\simeq 3$.

We have shown that the SKA will allow us to get information, not affected by
dust extinction,  on galaxy SFRs down to tens of $M_\odot/\hbox{yr}$ up to
the highest redshifts, thus extending by up to 3 orders of magnitude the
high-$z$ SFR functions derived from \textit{Herschel} surveys.

\section*{Acknowledgements}
We gratefully acknowledge many constructive comments by an anonymous referee,
that helped us improving this paper. Work supported in part by ASI/INAF
Agreement 2014-024-R.0 for the {\it Planck} LFI activity of Phase E2 and by
PRIN INAF 2012, project ``Looking into the dust-obscured phase of galaxy
formation through cosmic zoom lenses in the Herschel Astrophysical Large Area
Survey''.


\end{document}